\documentclass[11pt,a4paper,oneside]{article}
\usepackage{listings}             
\usepackage{color}
\usepackage{caption}
\usepackage{framed}
\usepackage{fancyhdr} 
\usepackage{lastpage} 
\usepackage{extramarks} 
\usepackage[margin=1in,tmargin=0.5cm]{geometry}
\usepackage{multirow}
\usepackage{fixltx2e}
\usepackage{multicol}
\usepackage{tipa}
\usepackage{amsmath}
\usepackage{amsfonts}
\usepackage{graphicx}
\usepackage{float}
\usepackage{subcaption}
\usepackage{chngcntr}
\usepackage[export]{adjustbox}
\usepackage[affil-it]{authblk}

\DeclareCaptionFont{white}{\color{white}}
\DeclareCaptionFormat{listing}{\colorbox[cmyk]{0.43, 0.35, 0.35,0.01}{\parbox{\textwidth}{\hspace{15pt}#1#2#3}}}

\newcommand{\hmwkTitle}{Identifying User Survival Types via Clustering of Censored Social Network Data} 

\newcommand{\hmwkClass}{}

\numberwithin{equation}{section}
\counterwithin{table}{section}
\counterwithin{figure}{section}

\begin{document}
\definecolor{mygreen}{rgb}{0,0.6,0}
\definecolor{mygray}{rgb}{0.5,0.5,0.5}
\definecolor{mymauve}{rgb}{0.58,0,0.82}

\title{
	\normalfont \small
	\textsc{\hmwkClass}\\	
	\rule{\linewidth}{0.5pt} 
	\Large
	 \textbf{\hmwkTitle}
	\rule{\linewidth}{1.5pt} 
}
	
\renewcommand*{\Authsep}{\qquad }
\renewcommand*{\Authand}{\qquad }
\renewcommand*{\Authands}{ \qquad}
\renewcommand*{\Affilfont}{\normalsize}

\author[1]{S Chandra Mouli}
\author[2]{Abhishek Naik}
\author[1]{Bruno Ribeiro}
\author[1]{Jennifer Neville}
\affil[1]{Purdue University}
\affil[2]{Indian Institute of Technology Madras}

\renewcommand\Affilfont{\itshape\small}

\date{}
\maketitle

\begin{abstract}
The goal of cluster analysis in survival data is to identify clusters that are decidedly associated with the survival outcome. Previous research has explored this problem primarily in the medical domain with relatively small datasets, but the need for such a clustering methodology could arise in other domains with large datasets, such as social networks. Concretely, we wish to identify different survival classes in a social network by clustering the users based on their lifespan in the network. In this paper, we propose a decision tree based algorithm that uses a global normalization of $p$-values to identify clusters with significantly different survival distributions. We evaluate the clusters from our model with the help of a simple survival prediction task and show that our model outperforms other competing methods.

\end{abstract}

\section{Introduction}
Survival analysis is used to model the length of time until a particular event occurs, for example, the time until death of a medical patient or failure of an equipment \cite{survivalanalysisbook}. One of the tasks in survival analysis is to cluster the individuals in a semi-supervised fashion by using not only their attributes but also their survival times. In other words, the goal of this task is to group the individuals with similar survival times into a single cluster. For example, the individuals could be divided into `high-risk' and `low-risk' groups, assuming there are only two clusters.

There are numerous applications for such a clustering procedure, one of which is the identification of cancer subtypes from gene expression data and is the most studied in the literature. Another example is that of grouping users based on their survival in a social network (i.e., the time until they leave the system permanently). Such an analysis can be extremely valuable as it can help in categorizing new users into groups that provide key information about their survival times. 

There have been previous attempts at an unsupervised approach to clustering where the clusters were identified by considering only the attributes and not the survival outcome \cite{uncluster1, uncluster2, uncluster3}. This approach has a clear drawback that the clusters obtained may be completely unrelated to the survival of these grouped individuals. A second approach to clustering is to divide the individuals purely based on their survival times \cite{deathcluster1, deathcluster2}. But these approaches do not provide us with any meaningful information about the connection between the features and the survival outcome. There also have been several semi-supervised clustering approaches \cite{supcluster1, supcluster2, supcluster3, supcluster4} proposed that use some form of label information to obtain the clusters, but most of these methods do not work well when the data is censored, which is a common characteristic of survival data. 

Only a few methods have been proposed that perform supervised clustering on censored data. Recently, Gaynor and Bair \cite{ssc} proposed a supervised version of the sparse clustering algorithm \cite{sparseclustering}. Sparse clustering provides a technique for feature selection in clustering by assigning weights to each feature. Supervised sparse clustering simply alters the initial weights of the features to reflect the features' relative importance in predicting survival. 

We note that the major expectation from the resultant clusters is that they have considerably different survival distributions (Section \ref{sub:surv}). In this paper, we utilize this fact and propose a novel partially supervised clustering approach for survival data. We predominantly work with social network data and obtain clusters of users based on their survival in the system.

\section{Preliminaries} \label{sec:conceptsinsurvivalanalysis}
In what follows, we describe the important concepts relating to survival analysis that are used in this paper.
\subsection{Survival Distribution \& Hazard Function} \label{sub:surv}
Survival distribution is defined as the probability that an individual survives atleast until time $t$, and is given by
\begin{align}
S(t) = P(T > t) = 1 - F(t), \qquad 0 < t < \infty,
\end{align}
where T is a nonnegative random variable representing the time of death of an individual, and $F(t)$ is the cumulative distribution function. 
In survival applications, it is typically convenient to define the hazard function, that represents the instantaneous rate of death of an individual given that she has survived till time $t$. The hazard function $\lambda(t)$, is given by
\begin{align}
\lambda(t) = \frac{f(t)}{S(t)},
\end{align}
where $f(t)$ is the probability density function. 
 
\subsection{Right Censoring}\label{sub:censor}
\begin{figure} 
\centering
\includegraphics[scale=0.4]{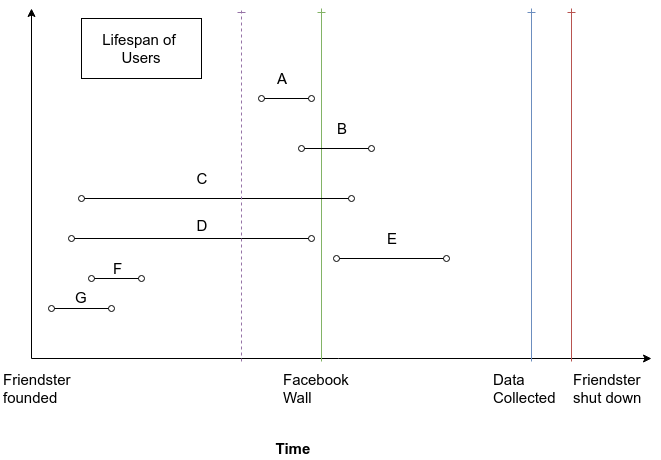}
\caption{Image depicting the lifespan of users in Friendster when comments are used as a proxy for activity. The vertical dotted line indicates the cut-off period of ten months. Users with no activity in this period are considered `dead'. In this figure, users A, B, C \& D have censored observations, user E is discarded and users F \& G have known survival lifetime.}
\label{fig:chart3}
\end{figure}

When working with survival times of individuals, it is common to have censored observations. This happens when the event in consideration does not occur until end of the study. Consider, for example, the time until a user in a social network stops being active. In this scenario, we say the observation of a user is (right) censored if, at the time of data collection, the user is still active, i.e., the death event has not yet occurred. It is clear that ignoring the effect of censoring can lead to skewed estimates of survival probabilities. Right censoring can be classified into three types, namely, Type-I, Type-II and Random censoring \cite{survivalanalysisbook}. Random censoring is a common feature when the individuals enter the study at different times, which is notably the case in the social network scenario, where the users join the system at different times (Figure \ref{fig:chart3}). Here, we assume that the censoring times are independent of the death times, which is justified when the joining times are random \cite{survivalanalysisbook}. In the following subsection, we define Kaplan-Meier estimator (or product limit estimator) that provides a method for incorporating the censoring effect while obtaining the survival probabilities. 

\subsection{Kaplan-Meier Estimator}\label{sub:kmest}
Kaplan-Meier estimator \cite{kaplanmeier} has been widely used in a variety of survival analysis tasks since its introduction. It provides a non-parametric maximum likelihood estimate of the empirical survival distribution, given by, 
\begin{align}
\hat{F}(t) = \prod_{j|t_j \leq t} \frac{n_j - d_j}{n_j},
\end{align}
where $d_j$ is the number of individuals who `die' at time $t_j$ and $n_j$ is the number of individuals at risk of `death' at time just prior to $t_j$, i.e., the individuals that are not `dead' and not yet been censored.

\section{Related Work}
Majority of work in survival analysis has dealt with the task of predicting the survival outcome especially when the number of features is much higher than the number of subjects \cite{predictsurvival1, predictsurvival2, predictsurvival3, predictsurvival4}. A number of approaches have also been proposed to perform feature selection in survival data \cite{featureselection1, featureselection2}. In the social network scenario, Sun et al. \cite{whenwillithappen} tried to predict the relationship building time, that is, the time until a particular link is formed in the network. They use generalized linear models \cite{glm} with a modified likelihood function that incorporates censoring.

There have been relatively fewer works that perform clustering on survival data. Many unsupervised approaches have been proposed to identify cancer subtypes in gene expression data \cite{uncluster1, uncluster2, uncluster3}. However, we are interested in the task of supervised clustering for survival data. Traditional semi-supervised clustering methods \cite{supcluster1, supcluster2, supcluster3, supcluster4} do not perform well in this scenario since they do not provide a way to handle the issues with right censoring. Bair and Tibshirani \cite{bair} proposed a semi-supervised method for clustering survival data in which they assign Cox scores \cite{coxproportionalhazardsmodel} for each feature (or gene) in their dataset and considered only the features with scores above a predetermined threshold. Then, an unsupervised clustering algorithm, like k-means, is used to group the individuals using only the selected features. Such an approach can miss out on clusters when they are weakly associated with the survival outcome since such features are discarded immediately after the initial screening.

In order to overcome this issue, Gaynor and Bair \cite{ssc} proposed supervised sparse clustering as a modification to the sparse clustering algorithm of Witten and Tibshirani \cite{sparseclustering}. The sparse clustering algorithm uses an objective function similar to k-means but with the modification that each feature has a weight associated to it. Supervised sparse clustering \cite{ssc} initializes these feature weights depending on the feature's relation with the survival outcome and optimizes the same objective function. Once again, they use Cox scores \cite{coxproportionalhazardsmodel} to quantify the effect of each feature on the survival outcome. The authors show that this leads to a clustering that is relatively more linked to the survival outcome. 

Both of these methods have been shown to perform well when the dataset size is small. Supervised sparse clustering in particular, is computationally expensive since in each iteration, it performs an unsupervised k-means clustering over the entire dataset. In this paper, we propose a decision tree based clustering algorithm that not only identifies better clusters than the existing methods but can also work efficiently with large amounts of data.

\section{Methodology}
Our primary goal for clustering is that the survival distributions be different across clusters. In this section, we present a decision tree based approach that is built to optimize for this goal.
The principal idea is to construct a decision tree such that the survival distributions of the two populations of users at each split differ significantly from each other. Concretely, we split the current set of users based on an attribute-value test and obtain the survival distributions of the two populations of users using Kaplan-Meier estimates (Section \ref{sub:kmest}).  We use Kuiper statistic \cite{kuipertest} in order to quantify how significantly these survival distributions differ. This process is repeated for all attribute-value pairs and the one that results in the lowest p-value (denoting that the survival distributions after the split are most likely different from each other) is used as a node in the decision tree. The significance level, $\alpha$, is a parameter to our algorithm (set at 0.05  in our experiments). It is important to note that we are performing many statistical tests at each node which leads to a multiple hypothesis test problem \cite{multiplehypothesistesting1, multiplehypothesistesting2}. We use the Bonferroni correction \cite{multiplehypothesistesting1} to compensate for doing $m$ statistical tests which reduces the significance level by a factor of $m$. Thus, a node is split only if the resultant p-value is below the corrected significance level $\alpha/m$.

This procedure results in a tree where each leaf node has an associated population of users and thus, the leaf nodes themselves can be interpreted as clusters. But, the issue here is that the leaf nodes need not have significantly different distributions from each other. It is not hard to imagine that two leaf nodes descending from different parts of the tree may have very similar survival distributions. Hence, it is necessary to group these leaf nodes such that the ones with similar distributions are clustered together. Note that growing a tree deep and clustering the leaf nodes is different from growing a shallow tree and using the leaf nodes as clusters.

\begin{figure}
\includegraphics[scale=0.8, right]{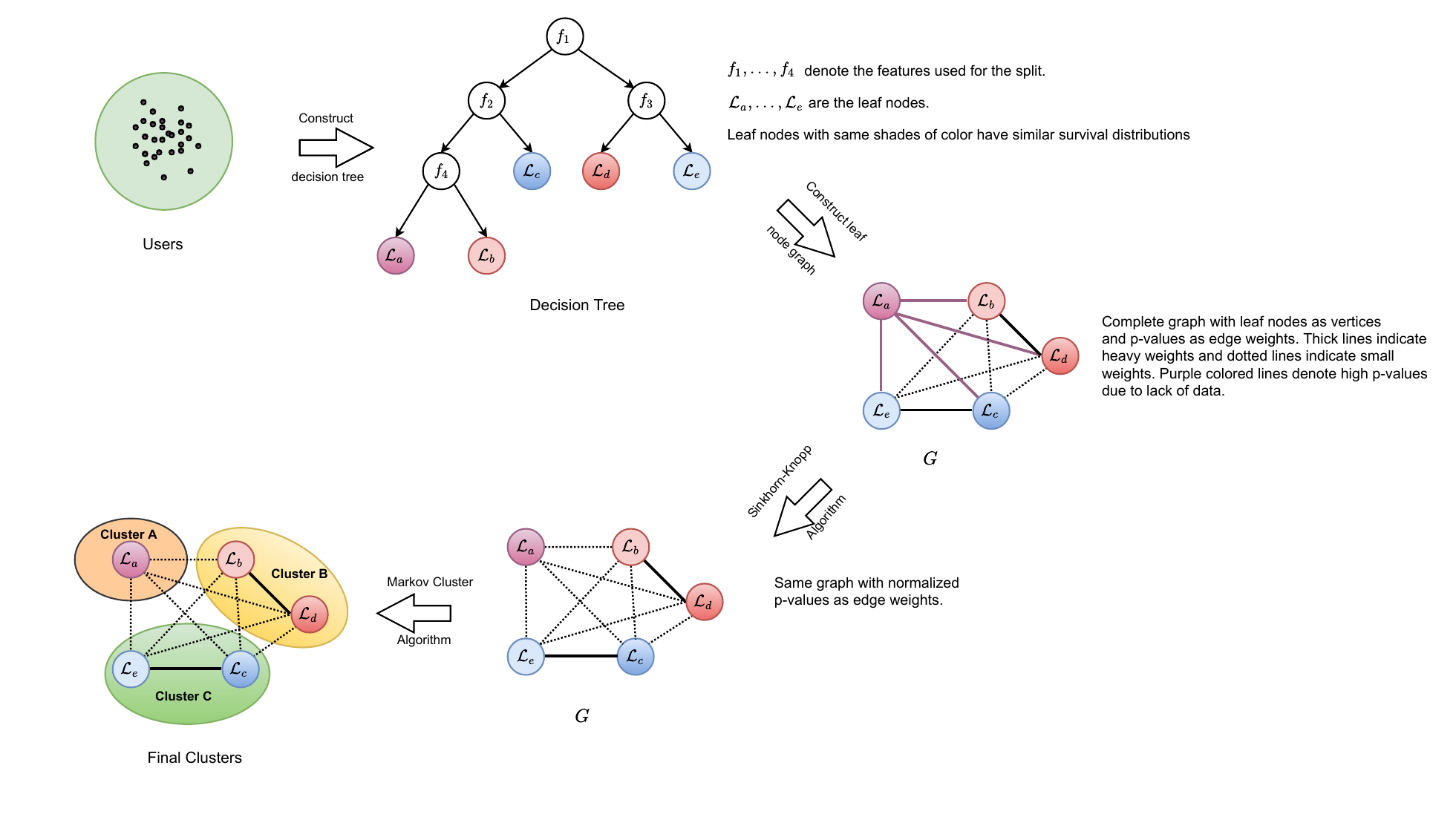}
\caption{Image depicting the complete procedure to obtain the clusters.}
\end{figure}

Let $\mathcal{L}$ be the set of leaf nodes. In order to cluster these leaf nodes, we build a complete graph $G = (V, E)$, where $V = \mathcal{L}$ and $E = \{(i, j) : i, j \in \mathcal{L}\}$. Define a weighted adjacency matrix $W \in \mathbb{R}_{\geq 0}^{|V| \times |V|}$ such that $W_{ij} = K_p(i, j)$ where $K_p : \mathcal{L} \times \mathcal{L} \rightarrow \mathbb{R}$ is a function that, given two leaf nodes, returns the p-value associated to the Kuiper's test between the survival distributions of the two leaf nodes. In other words, the weights on the edge represent the degree of similarity between the survival distribution of the two vertices. Now we could perform a graph clustering procedure on $G$ in order to cluster the leaf nodes.  However, using p-values directly as edge weights is not a sound approach. This is because p-values can be high for two reasons -- the distributions in question are very similar, or there is not enough data to significantly claim that the distributions differ from each other. Thus, a leaf node with very few associated users will form heavy edges with all the other leaf nodes, resulting in a clustering with just one group. We normalize the weight matrix $W$ using Sinkhorn-Knopp algorithm \cite{sinkhornknopp} that converts it into a doubly stochastic matrix $W_{sk}$, thereby solving the aforementioned issue. We use Markov cluster algorithm \cite{mcl} on the graph $G$ and weight matrix $W_{sk}$ in order to obtain a clustering of leaf nodes and consequently, a clustering on the entire set of users.

\section{Dataset}
In this paper, we analyze a large-scale social network dataset collected from Friendster. Friendster was founded in 2002 and was one of the earliest social networking websites, reaching 3 million users within the first few months \cite{friendsterstats}. The website allowed users to share messages, photos and videos with other members. Each user also had a profile page consisting of general information like name, gender, age, location and interests.

After processing 30TB of data, originally collected by the Internet Archive in June 2011, the resulting network has around 15 million users with 335 million friendship links. Each user has profile information such as age, gender, and marital status. Additionally, there are user comments on each other's profile pages with timestamps that indicate activity in the site. See Table 6.1 for some additional statistics on the dataset. 

\begin{table}[h!]
\centering
\begin{tabular}{|c|c|}
\hline
Number of Users & 15M\\
\hline
Number of friendship links & 335M\\
\hline
Number of comments & 75M\\
\hline
Number of users with atleast one comment & 9.5M\\
\hline
Number of users with atleast ten comments & 1.93M\\
\hline
Number of users with Age, Gender \& Location specified & 6.47M\\
\hline
\end{tabular}
\caption{Statistics on the Friendster dataset}
\end{table}

Since, we do not have users' login information, we use the comments sent and received by the users as a proxy for activity. We choose ten months of inactivity to be the cut-off period after which the user will be assumed to have left the social network. The time from the user's joining to her last comment will be considered as her lifetime in the system.

\begin{figure}
\centering
\includegraphics[scale=1]{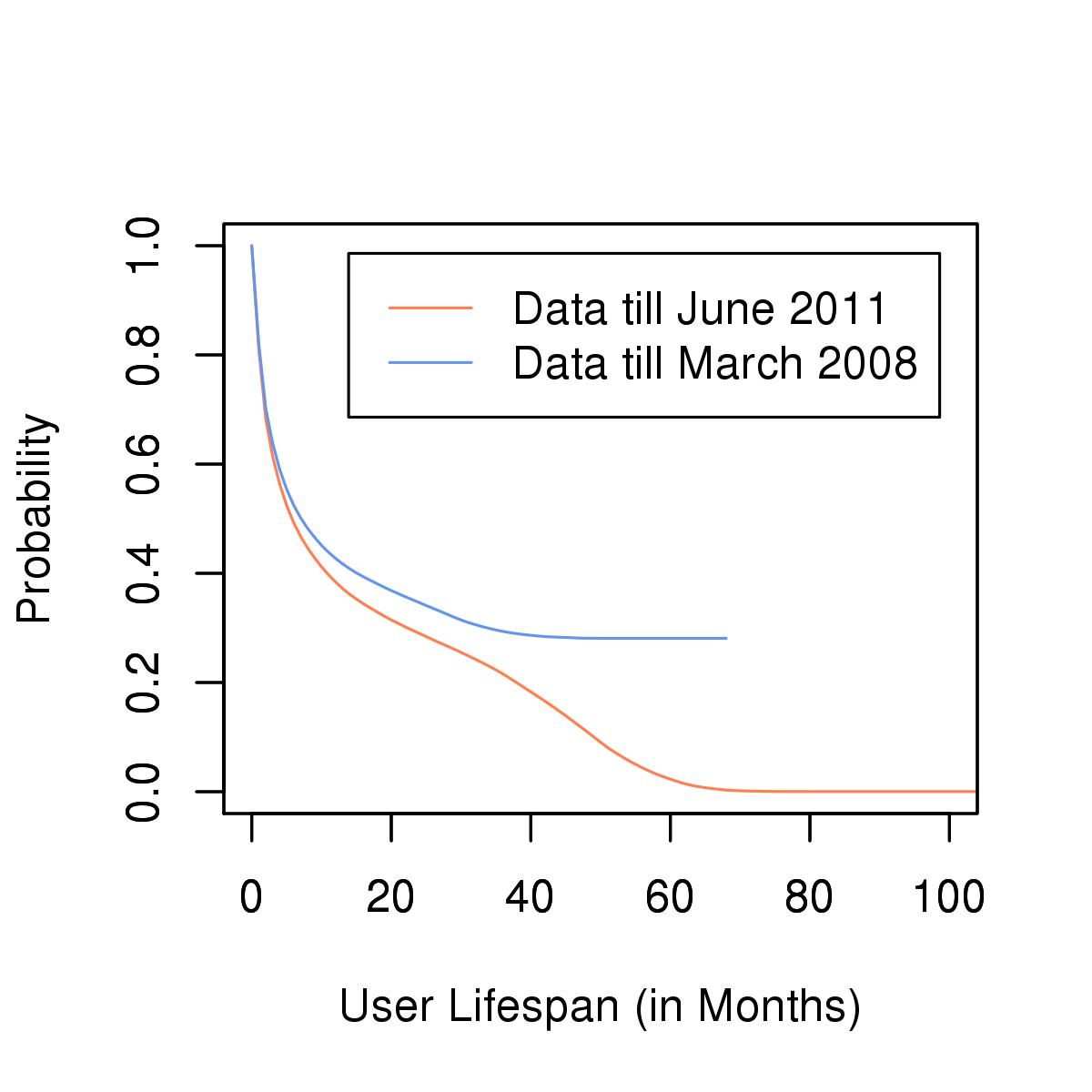}
\caption{Survival distributions of the complete data and the reduced data.}
\label{fig:kmplots}
\end{figure}

Ribeiro and Faloutsos \cite{brunowsdm} depicted the effect of the introduction of ``new Facebook wall'' in July 2008 to other competing social networking websites including Friendster. It is clear from their analysis that Friendster faced a continuous decline in the number of daily active users since then. Seeing that we wish to analyze the system on its own without any external influence, we only use the data upto March 2008 (six years from the introduction of Friendster) and disregard the rest. Figure \ref{fig:kmplots} shows the estimated survival distributions for the entire data and the reduced data. Note the sudden drop in survival probabilities when using the complete data, which is missing when we use only the data prior to the introduction of ``new Facebook wall''. In this work, we only consider a subset of 1.2 million users who had participated in atleast one comment, had specified their age and gender, and had joined the social network before March 2008. Our processed data will be made available to the public once we get the RIB approval for its distribution.

\section{Evaluation Metrics}
In this section, we describe a survival prediction task that we designed to evaluate the quality of the clusters obtained from our procedure.
We also describe briefly, two other standard evaluation techniques used in the literature, namely, hazard ratio and log rank test \cite{hazardratiousage, logranktest1}. 

\subsection*{Classification task} \label{sub:classification}
In order to validate our claim that the clusters obtained differentiate  the users based on their survival outcome, we devise a classification task as follows - \textit{given a user's profile and activity information for the initial five months, predict whether she will stay in the system five months hence}.

We obtain the clusters from running different clustering procedures on the features generated from the initial five months' data. We then use only these cluster labels as features in a logistic regression model \cite{glm} to predict whether the user will survive the next five months. A high prediction accuracy indicates that the clustering has extracted the information about the survival outcome from the entire set of features. 

\subsection*{Hazard ratio}
Hazard ratio is defined as the ratio of the hazard rates (Section \ref{sub:surv}) of two groups of entities. The Cox proportional hazards model \cite{coxproportionalhazardsmodel} provides a method to estimate the hazard ratio given that the hazard ratio is constant over time. Spruance et al. \cite{hazardratiousage} give a description of the interpretation and the correct usage of the hazard ratios. 

\subsection*{Log-rank test}
Log-rank test \cite{logranktest1, logranktest2} is a non-parametric hypothesis test that is widely used to compare two survival distributions. It tests the null hypothesis that the two (or more) groups in consideration have the same survival distributions. The predominant reason for the popularity of this test while comparing survival distributions is that it incorporates the effect of censoring the same way as the Kaplan-Meier estimates \cite{logranktestassumption}.

\section{Results}
We compare our model with two other clustering approaches -- semi-supervised clustering of Bair and Tibshirani \cite{bair}, and supervised sparse clustering by Gaynor and Bair \cite{ssc}. 
We use user's profile features (like age, gender, relationship status, occupation, location) as well as construct features based on the user's initial five months' activity (like number of comments sent and received, number of individuals interacted with, etc.). Since our model is based on decision trees, it can handle both categorical and numerical features with ease. In order for other clustering approaches to work effectively, we encode the categorical features like location using $g$ binary variables, where $g$ is the number of values the feature can take. Out of a total of 500 features, we choose 25 ($\approx \sqrt{500}$) top features found using Cox scores \cite{coxproportionalhazardsmodel}. Semi-supervised clustering uses only these top features to find the clusters. Supervised sparse clustering assigns positive weights to these features and zero weights to the rest, and runs the standard sparse clustering algorithm with these initial weights.

Table \ref{tab:tests} shows the values for the log-rank test and the hazard ratio for the clusters obtained from different clustering algorithms. The number of clusters were kept fixed at two in this experiment. The $\chi^2$ values shown in the table are huge, indicating that all three clustering algorithms return clusters that have significantly different distributions. The hazard ratios of clusters from our model and that from supervised sparse clustering are comparable. 

The performance of logistic regression model using only the cluster labels as features is presented in Table \ref{tab:classification} for the three clustering algorithms. We repeat the task for different values for $k$, the number of clusters . The clusters from our model have higher prediction accuracy than clusters from other models for $k = 2$ and $4$ whereas the accuracy is comparable for $k = 3$ and $5$. Our method also has higher f-measure scores compared to the competing models regardless of $k$.

\begin{table}[]
\centering
\begin{tabular}{|l|c|c|c|c|c|c|}
\hline
Clustering Method & Log Rank Test ($\chi^2$) & Hazard Ratio \\
\hline
Proposed Method & 172557 & 3.242\\
Semi-Supervised Clustering \cite{bair} & 141206 & 2.274\\
Supervised Sparse Clustering \cite{ssc} & 140660 & 3.331\\
\hline
\end{tabular}
\caption{Log Rank test and Hazard ratio values for $k$ = 2}
\label{tab:tests}
\end{table}

\begin{table}[]
\centering
\begin{tabular}{|c|c|c|c|c|c|}
\hline
& Precision  & Recall  & F-measure  & Accuracy  & FPR  \\
\hline
Proposed Method ($k$ = 2) & 0.689 & 	0.707 &	0.698&	0.673&	0.366 \\ 
Proposed Method ($k$ = 3) &0.711&	0.612&	0.658&	0.659&	0.286 \\
Proposed Method ($k$ = 4) & 0.688&	0.658&0.673&	0.657&	0.343\\
Proposed Method ($k$ = 5) & 0.689&	0.665&	0.677&	0.660&	0.345\\
\hline

Semi-Supervised Clustering ($k$ = 2) &0.763&	0.502&	0.605&	0.650&	0.178\\

Semi-Supervised Clustering ($k$ = 3) & 0.730&	0.584&	0.649&	0.662&	0.249\\
Semi-Supervised Clustering ($k$ = 4) & 0.511&	0.822&	0.630&	0.484&	0.906\\
Semi-Supervised Clustering ($k$ = 5) & 0.727&	0.591&	0.652&	0.662&	0.255\\
\hline

Supervised Sparse Clustering ($k$ = 2) & 0.764&	0.499&	0.604&	0.649&	0.176\\
Supervised Sparse Clustering ($k$ = 3) & 0.732&	0.579&	0.647&	0.661&	0.243\\
Supervised Sparse Clustering ($k$ = 4) & 0.772&	0.479&	0.591&	0.645&	0.162\\
Supervised Sparse Clustering ($k$ = 5) & 0.725&	0.509&	0.598&	0.633&	0.222\\
\hline
\end{tabular}
\caption{Classification results with features from various clustering algorithms for number of clusters, $k = 2,3,4,5$. The clusters obtained from the proposed method achieve better accuracies and f-scores when $k = 2$ and $4$ whereas the accuracies are comparable for $k = 3$. Highest accuracy across all algorithms is achieved by the proposed method when $k = 2$, that is, when there are only two classes of Friendster users: short-lived and long-lived.}
\label{tab:classification}
\end{table}

\section{Conclusion}
In this paper, we proposed a partially supervised approach for clustering users based on their survival outcome. We used decision trees to divide the users such that the survival distributions of the subgroups are significantly different at each step. We then performed graph clustering over these subgroups in order to make sure that the subgroups with similar survival distributions are clustered together. Explicitly working with survival distributions effectively leads to a clustering that is highly associated with the survival outcome. We used our model in a social network dataset to identify groups of users with different survival types. We evaluated our model using two standard metrics, log-rank test and hazard ratio, and a classification task that we devised to measure the clusters' ability to predict survival. We also observed in our dataset that the classification accuracy is highest when we use the proposed method to cluster the users into two groups -- \textit{short-lived} and \textit{long-lived}.

\bibliography{identifying_survival_types}

\begin{thebibliography}{10}

\bibitem{supcluster1}
Charu~C Aggarwal, Stephen~C Gates, and Philip~S Yu.
\newblock On using partial supervision for text categorization.
\newblock {\em IEEE Transactions on Knowledge and data Engineering},
  16(2):245--255, 2004.

\bibitem{uncluster2}
Ash~A Alizadeh, Michael~B Eisen, R~Eric Davis, Chi Ma, Izidore~S Lossos,
  Andreas Rosenwald, Jennifer~C Boldrick, Hajeer Sabet, Truc Tran, Xin Yu,
  et~al.
\newblock Distinct types of diffuse large b-cell lymphoma identified by gene
  expression profiling.
\newblock {\em Nature}, 403(6769):503--511, 2000.

\bibitem{bair}
Eric Bair and Robert Tibshirani.
\newblock Semi-supervised methods to predict patient survival from gene
  expression data.
\newblock {\em PLoS Biol}, 2(4):e108, 2004.

\bibitem{supcluster2}
Sugato Basu, Arindam Banerjee, and Raymond Mooney.
\newblock Semi-supervised clustering by seeding.
\newblock In {\em In Proceedings of 19th International Conference on Machine
  Learning (ICML-2002)}. Citeseer, 2002.

\bibitem{supcluster3}
Sugato Basu, Mikhail Bilenko, and Raymond~J Mooney.
\newblock A probabilistic framework for semi-supervised clustering.
\newblock In {\em Proceedings of the tenth ACM SIGKDD international conference
  on Knowledge discovery and data mining}, pages 59--68. ACM, 2004.

\bibitem{uncluster3}
Arindam Bhattacharjee, William~G Richards, Jane Staunton, Cheng Li, Stefano
  Monti, Priya Vasa, Christine Ladd, Javad Beheshti, Raphael Bueno, Michael
  Gillette, et~al.
\newblock Classification of human lung carcinomas by mrna expression profiling
  reveals distinct adenocarcinoma subclasses.
\newblock {\em Proceedings of the National Academy of Sciences},
  98(24):13790--13795, 2001.

\bibitem{logranktestassumption}
J~Martin Bland and Douglas~G Altman.
\newblock The logrank test.
\newblock {\em Bmj}, 328(7447):1073, 2004.

\bibitem{predictsurvival2}
Hege~M B{\o}velstad, St{\aa}le Nyg{\aa}rd, Hege~L St{\o}rvold, Magne Aldrin,
  {\O}rnulf Borgan, Arnoldo Frigessi, and Ole~Christian Lingj{\ae}rde.
\newblock Predicting survival from microarray data—a comparative study.
\newblock {\em Bioinformatics}, 23(16):2080--2087, 2007.

\bibitem{coxproportionalhazardsmodel}
David~R Cox.
\newblock Regression models and life-tables.
\newblock In {\em Breakthroughs in statistics}, pages 527--541. Springer, 1992.

\bibitem{uncluster1}
Michael~B Eisen, Paul~T Spellman, Patrick~O Brown, and David Botstein.
\newblock Cluster analysis and display of genome-wide expression patterns.
\newblock {\em Proceedings of the National Academy of Sciences},
  95(25):14863--14868, 1998.

\bibitem{ssc}
Sheila Gaynor and Eric Bair.
\newblock Identification of biologically relevant subtypes via preweighted
  sparse clustering.
\newblock {\em Biostatistics}, pages 1--33, 2013.

\bibitem{predictsurvival3}
Torsten Hothorn, Peter B{\"u}hlmann, Sandrine Dudoit, Annette Molinaro, and
  Mark~J Van Der~Laan.
\newblock Survival ensembles.
\newblock {\em Biostatistics}, 7(3):355--373, 2006.

\bibitem{featureselection1}
Hemant Ishwaran, Udaya~B Kogalur, Eiran~Z Gorodeski, Andy~J Minn, and Michael~S
  Lauer.
\newblock High-dimensional variable selection for survival data.
\newblock {\em Journal of the American Statistical Association},
  105(489):205--217, 2010.

\bibitem{kaplanmeier}
Edward~L Kaplan and Paul Meier.
\newblock Nonparametric estimation from incomplete observations.
\newblock {\em Journal of the American statistical association},
  53(282):457--481, 1958.

\bibitem{kuipertest}
Nicolaas~H Kuiper.
\newblock Tests concerning random points on a circle.
\newblock In {\em Indagationes Mathematicae (Proceedings)}, volume~63, pages
  38--47. Elsevier, 1960.

\bibitem{featureselection2}
Vincenzo Lagani and Ioannis Tsamardinos.
\newblock Structure-based variable selection for survival data.
\newblock {\em Bioinformatics}, 26(15):1887--1894, 2010.

\bibitem{logranktest1}
Nathan Mantel.
\newblock Evaluation of survival data and two new rank order statistics arising
  in its consideration.
\newblock {\em Cancer chemotherapy reports. Part 1}, 50(3):163--170, 1966.

\bibitem{glm}
Peter McCullagh.
\newblock Generalized linear models.
\newblock {\em European Journal of Operational Research}, 16(3):285--292, 1984.

\bibitem{survivalanalysisbook}
Rupert~G Miller~Jr.
\newblock {\em Survival analysis}, volume~66.
\newblock John Wiley \& Sons, 2011.

\bibitem{supcluster4}
Kamal Nigam, Andrew McCallum, Sebastian Thrun, Tom Mitchell, et~al.
\newblock Learning to classify text from labeled and unlabeled documents.
\newblock {\em AAAI/IAAI}, 792, 1998.

\bibitem{logranktest2}
Richard Peto and Julian Peto.
\newblock Asymptotically efficient rank invariant test procedures.
\newblock {\em Journal of the Royal Statistical Society. Series A (General)},
  pages 185--207, 1972.

\bibitem{brunowsdm}
Bruno Ribeiro and Christos Faloutsos.
\newblock Modeling website popularity competition in the attention-activity
  marketplace.
\newblock In {\em Proceedings of the Eighth ACM International Conference on Web
  Search and Data Mining}, pages 389--398. ACM, 2015.

\bibitem{friendsterstats}
Gary Rivlin.
\newblock Wallflower at the web party.
\newblock {\em New York Times}, 15(10), 2006.

\bibitem{multiplehypothesistesting1}
G~Rupert~Jr et~al.
\newblock {\em Simultaneous statistical inference}.
\newblock Springer Science \& Business Media, 2012.

\bibitem{multiplehypothesistesting2}
Juliet~Popper Shaffer.
\newblock Multiple hypothesis testing.
\newblock {\em Annual review of psychology}, 46(1):561--584, 1995.

\bibitem{deathcluster1}
Margaret~A Shipp, Ken~N Ross, Pablo Tamayo, Andrew~P Weng, Jeffery~L Kutok,
  Ricardo~CT Aguiar, Michelle Gaasenbeek, Michael Angelo, Michael Reich,
  Geraldine~S Pinkus, et~al.
\newblock Diffuse large b-cell lymphoma outcome prediction by gene-expression
  profiling and supervised machine learning.
\newblock {\em Nature medicine}, 8(1):68--74, 2002.

\bibitem{predictsurvival4}
Pannagadatta~K Shivaswamy, Wei Chu, and Martin Jansche.
\newblock A support vector approach to censored targets.
\newblock In {\em Data Mining, 2007. ICDM 2007. Seventh IEEE International
  Conference on}, pages 655--660. IEEE, 2007.

\bibitem{sinkhornknopp}
Richard Sinkhorn and Paul Knopp.
\newblock Concerning nonnegative matrices and doubly stochastic matrices.
\newblock {\em Pacific Journal of Mathematics}, 21(2):343--348, 1967.

\bibitem{hazardratiousage}
Spotswood~L Spruance, Julia~E Reid, Michael Grace, and Matthew Samore.
\newblock Hazard ratio in clinical trials.
\newblock {\em Antimicrobial agents and chemotherapy}, 48(8):2787--2792, 2004.

\bibitem{whenwillithappen}
Yizhou Sun, Jiawei Han, Charu~C Aggarwal, and Nitesh~V Chawla.
\newblock When will it happen?: relationship prediction in heterogeneous
  information networks.
\newblock In {\em Proceedings of the fifth ACM international conference on Web
  search and data mining}, pages 663--672. ACM, 2012.

\bibitem{deathcluster2}
Marc~J Van De~Vijver, Yudong~D He, Laura~J Van't~Veer, Hongyue Dai,
  Augustinus~AM Hart, Dorien~W Voskuil, George~J Schreiber, Johannes~L Peterse,
  Chris Roberts, Matthew~J Marton, et~al.
\newblock A gene-expression signature as a predictor of survival in breast
  cancer.
\newblock {\em New England Journal of Medicine}, 347(25):1999--2009, 2002.

\bibitem{mcl}
Stijn~Marinus Van~Dongen.
\newblock {\em Graph clustering by flow simulation}.
\newblock PhD thesis, 2001.

\bibitem{sparseclustering}
Daniela~M Witten and Robert Tibshirani.
\newblock A framework for feature selection in clustering.
\newblock {\em Journal of the American Statistical Association},
  105(490):713--726, 2010.

\bibitem{predictsurvival1}
Daniela~M Witten and Robert Tibshirani.
\newblock Survival analysis with high-dimensional covariates.
\newblock {\em Statistical methods in medical research}, 19(1):29--51, 2010.

\end{thebibliography}
\bibliographystyle{plain}

\end{document}